\documentclass{aa}  
\bibpunct{(}{)}{;}{a}{}{,}

\usepackage{afterpage,natbib,lipsum}
\usepackage{graphicx}
\usepackage{txfonts}
\begin{document}

   \title{
   Cosmic expansion history from SNe Ia data\\ via information field theory -- the \texttt{charm} code}


   \author{Nat\`{a}lia Porqueres\inst{1} \and Torsten A. En{\ss}lin\inst{2} \and Maksim	Greiner\inst{2} \and Vanessa 	B\"{o}hm\inst{2} \and  Sebastian Dorn\inst{2} \and Pilar Ruiz-Lapuente\inst{4,3} \and Alberto Manrique\inst{1,3}}

   \institute{University of Barcelona, Departament de F\'{i}sica Qu\`{a}ntica i Astrof\'{i}sica, Mart\'{i} i Franqu\`{e}s 1, 08028 Barcelona, Spain
         \and                     
             Max-Planck-Insitut f\"{u}r Astrophysik (MPA),
             Karl-Schwarzschild-Strasse 1 , D-85741 Garching, Germany
         \and 
             Institut de Ci\`{e}ncies del Cosmos,  Mart\'{i} i Franqu\`{e}s 1, 08028 Barcelona, Spain
         \and 
             Instituto de F\'{i}sica Fundamental, CSIC, Serrano 121, 28006 Madrid, Spain  \\
             }

   \date{Received 13/08/2016; accepted 03/12/2016}

 
  \abstract
   {We present  \texttt{charm} (cosmic history agnostic reconstruction method), a novel inference algorithm that reconstructs the cosmic expansion history as encoded in the Hubble parameter $H(z)$ from SNe Ia data. The novelty of the approach lies in the usage of information field theory,
   a statistical field theory that is very well suited for the construction of optimal signal recovery algorithms. The \texttt{charm} algorithm infers non-parametrically $s(a)=\ln(\rho(a)/\rho_{\mathrm{crit}0})$, the density evolution which determines $H(z)$, without assuming an analytical form of $\rho(a)$ but only its smoothness with the scale factor $a=(1+z)^{-1}$. The inference problem of recovering the signal $s(a)$ from the data is formulated in a fully Bayesian way. In detail, we have rewritten the signal as the sum of a background cosmology and a perturbation. This allows us to determine the maximum a posteriory estimate of the signal by an iterative Wiener filter method. Applying  \texttt{charm} to the Union2.1 supernova compilation, we have recovered a cosmic expansion history that is fully  compatible with the standard $\Lambda$CDM cosmological expansion history with parameter values consistent with the results of the Planck mission.
   
   }


   \keywords{Cosmic expansion history --
                Information Field Theory --
                Supernova Ia 
               }

	\titlerunning{Cosmic expansion history from SNe Ia data -- the \texttt{charm} code}
	\authorrunning{Porqueres et al.}
   \maketitle 
%

\section{Introduction}

	   Combined observations of nearby and distant type Ia Supernovae (SNe Ia) have demonstrated that the expansion of the Universe is accelerating in the current epoch \citep{Perl1999, Rie1998}. Such a Universe can be described by the $\Lambda$CDM model, in which the cosmic acceleration is determined by Einstein's cosmological constant with a time-independent equation of state, $\omega \equiv p/\rho= -1$. However, this is just one of the possible explanations of the expansion that is consistent with the SNe Ia measurements. Others include a new field component filling the Universe as a quintessence or modified gravity \citep{Kas2015}. 
	   
	   Constraining the cosmic expansion as a function of redshift is a task of major interest, since the evolution of the scale factor allows us to probe properties of the fundamental components of the Universe. This may lead to a better understanding of their nature as well possibly providing evidence for new fundamental physics. 
	
	Recent studies of the baryonic acoustic oscillations (BAO) have suggested different constraints on the density of dark energy at high redshifts \citep{Del2015, Hee2016}. Such a change in the evolution of the dark energy density in the early epoch could be determined from SNe Ia data at high redshifts ($z>1$), which will be available shortly \citep{Rub2016}. In addition, some years from now a sample of $10^{5}$ SNe Ia is expected to be available from the LSST \citep{LSS2009}. This upcoming data will open an entirely new chapter in the study of dark energy.
	
	The aim of this work is to reconstruct the cosmic expansion history, encoded in the Hubble parameter $H(z)$, from supernovae data in the framework of Information Field Theory (IFT) \citep{Ens2009}. Conceptionally, IFT is a statistical field theory that permits the construction of optimal signal recovery algorithms. To this end, we developed the  \texttt{charm}\footnote{\texttt{charm} stands for cosmic history agnostic reconstruction method.} code, which is freely available.\footnote{https://gitlab.mpcdf.mpg.de/natalia/charm} We use the Union2.1 Supernova compilation, which is a database that contains 580 SNe Ia in the redshift range of $0.015<z<1.414$.
	
	Deriving the cosmic expansion history is a major goal of modern cosmology. To date, the low-redshift evolution of the Hubble parameter $H(z)$ has been studied with different methods. Some recent studies present analysis of the cosmic expansion by $\chi^2$ minimization \citep{Ber2016, Mel2015} while others develop non-parametric methods to solve the inverse problem of reconstructing the Hubble parameter $H(z)$ \citep{Li2016, Mon2014,  Ish2011, Arm2006} or the equation of state of dark energy \citep{Esp2005, Sim2005, Esp2008, Gen2009}. 
	Common to all non-parametric reconstructions, the ones cited above and the one we develop here, is that a quantity to be reconstructed (Hubble parameter, cosmic density, equation of state, etc.) as well a regularization for the otherwise ill-posed inference problem must be chosen. The discussed methods differ in what regularization is chose.   
	
	Here, we develop a non-parametric reconstruction in natural coordinates for the reconstruction of the logarithm of the cosmic density $s=\ln(\varrho/\varrho_{\mathrm{crit}0})$ as a function of the logarithm of the cosmic scale factor $x=-\ln a$ and thereby the Hubble parameter $H(z)$. The regularization arises from a Bayesian prior on potential solutions
$s(x)$. We construct this prior from the insight that constituents of the
cosmic density are likely to scale with the inverse scale factor to some power typically (but not exclusively) between zero (cosmological constant) and four (radiation). Translated to the log-log coordinates we advocate this to be natural, this means that straight lines in $s(x)$ are preferred over curved ones. We can also motivate the level of expected curvature: a transition from radiation to dark energy domination within a few e-folds of expansion has to be possible if our standard cosmological expansion history should be embraced by the prior.  
	
	An advantage of the adopted Bayesian methodology lies in the fact that it provides a flexible framework to question data: it can reconstruct the cosmic expansion history using different priors. For example, it can be asked how much the data requests a modification of a given cosmology or what the prefered expansion history is from a cosmological composition agnostic point of view. The main assumption is a smooth behavior of the logarithm of the density $\ln \rho$ with the logarithmic scale factor, $\ln a$, whereas the strength of this assumption can also be varied. 
	
	We probe that  \texttt{charm} is sensitive to features in expansion history at any low-redshift, $z<1.5$. In addition, the algorithm is easily extendible to include other datasets, such as BAO or Cepheids \citep{Rie2016}, which provide information of a transition epoch between deceleration and acceleration of the cosmic expansion \citep{Mor2016, Hee2016}.
	
	We develop and test  \texttt{charm}, so that it is ready for application to the new catalog Union3 compilation, which is expected to provide information about the dark energy density at high-redshifts. 
	
	After this introduction, we establish our notation and present our assumptions and the inference problem in Section 2. In Section 3, the SN Ia catalog is described and we derive our reconstruction method in Section 4. In Section 5, we specify our prior knowledge and the cosmological expansion histories that we use to test  \texttt{charm}. We present a comparison of  \texttt{charm} with previous literature in Section 6. Finally, we present the results of the reconstruction in Section 7 and conclude in Section 8.


\section{Inference approach}
\subsection{Basic notation}
	First of all, we needed to establish some notations and assumptions. We derived the algorithm of \texttt{charm} in the framework of IFT, following the notation of \cite{Ens2009}. We have assumed that we are analyzing a discrete set of data $d$ which may depend on a signal $s$, which contains the physical quantities of interest. In this case the signal is a field, $s(x)$, chosen to be the logarithm of the cosmic density $\rho(z)$ as a function of the logarithmic scale factor, $x=-\ln (a) = \ln (1+z)$, where $z$ is the redshift.
	
	This parametrization is natural, as it deals with dimensionless quantities, represents cosmic periods dominated by a constant equation of state $\rho \propto a^{1+\omega}$ as straight lines $s = s_0 + (1+\omega)x$, and converts relative variations of $\mathcal{O}(e)$ to additive variations of $\mathcal{O}(1)$. 
	
	This coordinate system  has the advantage that we can model the signal uncertainties by Gaussian fluctuations around a background cosmology, denoted as $t_\mathrm{bg}(x)$:
\begin{equation}	
\mathcal{P}(s) = \mathcal{G}(s-t_\mathrm{bg}, S) = \frac{1}{\sqrt{|2\pi S|}}\exp\Big(-\frac{1}{2}(s-t_\mathrm{bg})^\dagger S^{-1}(s-t_\mathrm{bg})\Big),
\label{Gaussian-field}
\end{equation}
where $S$ is the prior covariance matrix 
$S = \langle \phi\,\phi^{\dagger} \rangle_{(s|S)}$ with $\phi = s-t_\mathrm{bg}$.
Scalar products of continuous quantities are defined as $a^\dagger b \equiv \int^{\infty}_0 dx \, a(x)\, b(x)$.

The diagonal of the prior covariance, $S_{xx}= \langle \phi_x^2 \rangle_{(s|S)}$, 
encodes how much variation of the signal around the a priori
background cosmology $t_\mathrm{bg}$ is expected a priori at every location $x$. 
The off-diagonal terms 
of the covariance, $S_{xy}= \langle \phi_x \phi_y \rangle_{(s|S)}$, 
specify how correlated such deviations form the background cosmology are 
expected to be between the points $x$ and $y$. A larger correlation corresponds to smoother structures of the 
deviations. In Sect.~\ref{cosmo_prior}, we will use simple
and intuitive arguments about the expected roughness of $s$ to specify $S$, 
as well as different choices of the background cosmology $t_\mathrm{bg}$. In particular,
as no location of cosmic history is singled out a priori on a logarithmic scale, 
the prior covariance structure should be homogeneous,
$S_{xy}= C_s(|x-y|)$, with $C_s(r)$
a correlation function that only depends on the distance $r=|x-y|$.

\subsection{Signal inference}
	In the inference problem, we are interested in the probability of the signal given the data. This is described by the posterior $P(s|d)$, given by Bayes' Theorem,
\begin{equation}
P(s|d) = \frac{P(d|s)P(s)}{P(d)},
\end{equation}
which is the product of the likelihood $P(d|s)$ and the signal prior $P(s)$ normalized by the evidence $P(d)$. 

In the framework of IFT, inference problems are formulated in the language of statistical field theory. To that end we rewrite the posterior $P(s|d)$ as
\begin{equation}
P(s|d) = \frac{P(d|s)P(s)}{P(d)} = \frac{1}{\mathcal{Z}}e^{-\mathcal{H}(d,s)},
\end{equation}
where $\mathcal{H}$ is called the information Hamiltonian and $\mathcal{Z}$ is the partition function. They are defined as
\begin{eqnarray}
&& \mathcal{H}(d,s) = -\ln(P(d|s)P(s)), \\
&& \mathcal{Z}(d) = \int \mathcal{D}s \, e^{-\mathcal{H}(d,s)} = \int \mathcal{D}s \, P(d|s)P(s) = P(d),
\end{eqnarray}
where $\mathcal{D}s$ is a phase space integral.

	The information Hamiltonian comprises all available information and is for this reason the central mathematical object in our method development. Its minimum	with respect to $s$ for a given dataset $d$, for example, is the maximum of the posterior (MAP). Our algorithm calculates the MAP estimation \citep{Lem1999} of the expansion history. 
	

\section{Database}
	SNe Ia have been found to be an excellent probe for studying the expansion history of the Universe. Observations of the nearby and distant SNe Ia led to the discovery of the accelerating expansion \citep{Perl1999, Rie1998}. For this reason, we choose supernovae as our initial data set to develop an IFT based method for reconstructing the cosmic expansion history.

\begin{figure}[t]
\centering
\includegraphics[width=3.88in]{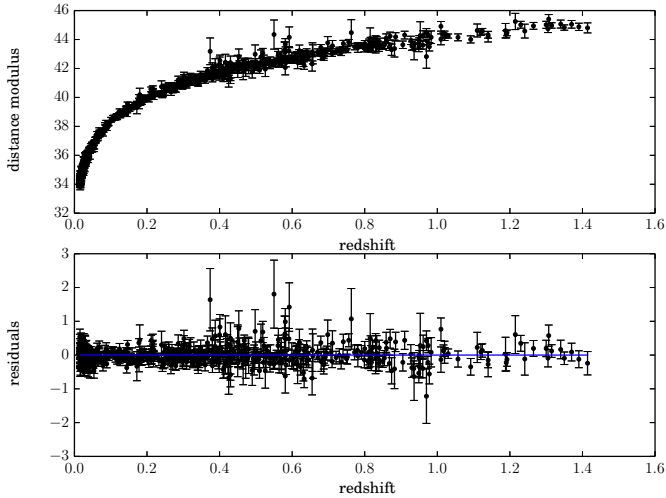}
\caption{Hubble diagram for the Union2.1 compilation (upper panel) and residuals of data respect to the Planck cosmology (bottom panel)}
\label{fig:data}
\end{figure}	
	
	\cite{Kow2008} described a method to analyze the SNe Ia data in a homogeneous manner and created a compilation, the Union SN Ia compilation, combining the world's SN data sets. Both new data and literature SNe were fit using a spectral template for lightcurve fitting, also known as SALT \citep{Guy2005}. 
	
	
	Here, we use the Union2.1 compilation, which contains 580 SNe Ia: the 557 data from Union2 and 23 new measurements at redshift $z>1$ \citep{Suz2012}. The data are distributed in a redshift range of $0.015<z<1.414$ corresponding to an x-range of $0.014 < x < 0.881$. Union2.1 provides the redshift, distance modulus and distance modulus error for each supernova, which are shown in Figure \ref{fig:data}. The catalog also includes uncertainty covariance matrix with systematics.

	In this work, we attempt a tomographic inversion of the SN data into a cosmic expansion history, where the term tomographic means a reconstruction of a searched object properties distributed along a coordinate for which the allocation does not follow from observation, in this case a reconstruction along the line of sight. Tomography is very sensitive to systematic errors and therefore, we should use a non-diagonal noise covariance matrix $N$ in our reconstruction in order to account for correlated systematic uncertainties. Some systematic errors are common in all the observations while other sources of systematic errors are controlled by the individual observers.	
	
	\cite{Kow2008}, identified two categories of systematic errors: the ones that affect measurements independently for each SN Ia (e.g. due to observational method) and systematic errors that affect the measurements of SNe at similar redshifts (e.g. due to astrophysics). Since the different sources of systematic errors can be considered to be independent, the total error can be well approximated as a Gaussian error. In Appendix \ref{appA} we briefly discuss the sources of systematics. 
	
	
	\section{Reconstruction}
	In order to reconstruct the cosmic expansion history, which is encoded in the Hubble parameter $H(z)$, from supernovae data we write $H$ in terms of the scale factor $a=(1+z)^{-1}$. Then we have
\begin{equation}
H=\frac{\dot{a}}{a}
\end{equation}
and
\begin{equation}
\left(\frac{H}{H_0}\right)^2=\frac{\rho}{\rho_{\mathrm{crit}0}},
\end{equation}
where $\rho_{\mathrm{crit}0}=3H_0^2/(8\pi G)$. The density $\rho$ is usually assumed to be polynomial in $a$ with three contributing terms: 
\begin{equation}
\rho(a)=\rho_{\Lambda}a^{0}+\rho_\mathrm{k}a^{-2}+\rho_\mathrm{DM}a^{-3}+\rho_\mathrm{rad}a^{-4}.
\end{equation}

	However, as we aim for a non-parametric reconstruction, we do not assume this polynomial form but just that $\rho$ is smooth with $a^{-1}$ as well as that power-law like building blocks of $\rho(a)$ are favored. 
	
	As data we use the distance moduli of SNe from Union2.1, assuming Gaussian noise. As our signal we define 
\begin{equation}
s(x)=\ln(\rho(x)/\rho_{\mathrm{crit}0}), 
\label{eq:sigdef}
\end{equation}
with
\begin{equation}
x=-\ln(a)=\ln(1+z).
\label{eq:corddef}
\end{equation}
The distance modulus is
\begin{equation}
\mu(z)=5 \log_{10}\Big[(1+z)d_H\int_0^z dz' \frac{1}{E(z')}\Big]-5,
\label{mu}
\end{equation}
where $d_H$ is constant in a flat universe, $d_H=c/H_0$, and
$E=H/H_0$. In $\Lambda$CDM we have
\begin{equation}
E(z)=\sqrt{\Omega_m(1+z)^3+\Omega_\Lambda},
\end{equation}
while our generic parametrization we have
\begin{equation}
 E(z) = \exp(s(x)/2).
\end{equation}

	\subsection{Response operator}
	
	Our cosmology signal $s$ has been imprinted onto the data. The functional relationship between the noiseless data $\mu$ and the signal can be regarded as a non-linear response operator. To be precise, the response operator describes how the distance moduli at different redshifts of the SNe depend on the unknown cosmic expansion history encoded in our signal $s$. 
	
%

Combining our parametrization (Eqs.~\eqref{eq:sigdef}~\&~\eqref{eq:corddef}) with eq.~\eqref{mu} yields the response as
\begin{equation}
R_j(s)=5 \log_{10}\Big[e^{x_j}d_H\int^{x_j}_0 dx' e^{-\frac{1}{2}s(x')+x'}\Big]-5,
\label{non-lin-response}
\end{equation}
with $x_j = \ln (1+z_j)$.


The response depends on an intergral over our signal $s(x)$. The goal of our tomographic method is to invert this integration. Since the response is not a linear operation of the form $R(s) = R\,s + \mathrm{const}$ it cannot be algebraically inverted, which would allow us to minimize the information Hamiltonian directly.
	
	
	\subsection{Expansion of the information Hamiltonian}
	In order to build the information Hamiltonian $\mathcal{H}(d,s) = -\ln P(d|s)-\ln P(s)$, we describe our a priori knowledge on $s$ with a Gaussian $P(s) = \mathcal{G}(s-t_\mathrm{bg},S)$ as described in eq. (\ref{Gaussian-field}). The background cosmology $t_\mathrm{bg}$ and the signal covariance $S$ are to be specified later. The noise is assumed to be Gaussian and independent of the signal, distributed as $P(n|s)=\mathcal{G}(n,N)$, since the sources of uncertainty which operate on the distance modulus $\mu$ can be approximately assumed to be independent from $\mu$.
	
	To perturbatively expand our problem, we can write the information Hamiltonian in terms of $s=t+\varphi$, where we call $t$ the pivot field and $\varphi$ its perturbation. By assuming a signal covariance $S$, the expanded Hamiltonian reads
\begin{eqnarray}
\mathcal{H}(d,\varphi|t)&=&-\log\frac{1}{|2\pi S|^{1/2} |2\pi N|^{1/2}}+ \nonumber\\
	&&+\frac{1}{2}\Big(d-R(t+\varphi)\Big)^\dagger N^{-1}\Big(d-R(t+\varphi)\Big)+ \nonumber\\
&&+\frac{1}{2}(\varphi+t-t_\mathrm{bg})^\dagger S^{-1} (\varphi+t-t_\mathrm{bg}),
\label{Hamiltonian-our-case}
\end{eqnarray}
where the prior background cosmology only affects the terms related to our prior. The amplitude of the perturbation is assumed to be $|\varphi|<1$ and then we can expand the response operator since $\exp(t+\varphi) \approx \exp(t) (1+\varphi)$.
	
	Our posterior is a non-Gaussian probability distribution function. In order to minimize the probability in an efficient and numerically robust way, we Gaussianize it by linearizing the response $R(s)$ around the pivot field $t$,
\begin{equation}
R(s) = R(t)+\frac{\partial R}{\partial\varphi}|_{s=t}\varphi+\mathcal{O}(\varphi^2) = R(t) + R_t \varphi + O(\varphi^2).
\end{equation}
The linearized response $R_t$ is 
\begin{equation}
 \left(R_t\right)_{jx} = -\alpha \frac{q_{jx}}{r_j},
\end{equation}
with $\alpha = 5/(2 \ln 10)$,
\begin{equation}
 q_{jx}\equiv d_H\, e^{-\frac{1}{2}t_x+x+x_j}\theta(x_j-x),
\end{equation}
and
\begin{equation}
 r_j \equiv e^{x_j}d_H\int^{x_j}_0 dx' e^{-\frac{1}{2}t_{x'}+x'}.
\end{equation}

	The resulting approximated Hamiltonian,
\begin{eqnarray}
\mathcal{H}(d,\varphi|t) &=& \frac{1}{2}\left(d-R(t)-R_t\varphi\right)^\dagger N^{-1}\left(d-R(t)-R_t\varphi\right) \\ \nonumber
&&+\frac{1}{2}(\varphi+t-t_\mathrm{bg})^\dagger S^{-1} (\varphi+t-t_\mathrm{bg}),
\end{eqnarray}
is then minimized by $m=Dj$ with 
\begin{equation}
D = (R_t^\dagger N^{-1}R_t + S^{-1})^{-1}, 
\end{equation}
the so-called information propagator operator and
\begin{equation}
j = R_t N^{-1} \left(d-R(t)\right) - S^{-1}\left(t-t_\mathrm{bg}\right),
\end{equation}
the so-called information source field.

The terms information source $j$ and information propagator $D$ may require a brief explanation. 
In this linear response approximation the Hamiltonian is quadratic in $s$. This corresponds to a Gaussian joint probability of signal and data, implying a Gaussian posterior,
\begin{equation}
 P(\varphi|d) \approx \mathcal{G}(\varphi-m, D).
\end{equation}
In this approximation the information propagator is the a posterior uncertainty covariance 
$D = \langle (\varphi-m)\,(\varphi-m)^{\dagger} \rangle_{(s|d,S)}$, 
very similar to the a priori covariance $S = \langle (s-t_\mathrm{bg}) \,(s-t_\mathrm{bg})^{\dagger} \rangle_{(s|S)}$.
The covariance $D$ is smaller than $S$, since the presence of the data has restricted the set 
of possible cosmologies. The information source $j$ contains the data and has excited the increased 
knowledge on $s$. This approximative linear solution to the inference problem is known as Wiener filter 
solution and $D$ is also called the Wiener variance.

	This Wiener filter solution is not at the minimum of the unapproximated Hamiltonian. To reach it we have to repeat the procedure after shifting the pivot cosmology to $t \leftarrow t+\varphi$ until $t$ no longer changes. We initialize the iteration by $t = t_\mathrm{bg}$ and stop it when $t$ becomes stationary. Our iterating Wiener filtering scheme can be regarded as a Newton method minimizing the full original Hamiltonian, regularized by expanding the response and not the Hamiltonian to ensure convergence of the Newton method, for which we had to simplify our propagator operator (Appendix \ref{apb}). Therefore, the correct MAP solution will be found at the end of the iterations. 

%
%
%

	
	\section{Cosmology prior\label{cosmo_prior}}
	
	In this section, we focus on the definition of the prior as a Gaussian process characterized by a power spectrum, which varies around a background cosmology \citep{Lem1999,Ens2009}. 
	
	First, we consider that the background expansion history $t_\mathrm{bg}$ is that of the $\Lambda$CDM model, which is the currently accepted cosmology. To validate our reconstruction, we take two more expansion histories: (1) A CDM cosmology, without a dark energy, in order to check that \texttt{charm} is able to recover the cosmological constant even if the prior does not favor its presence. (2) An agnostic cosmology that has little knowledge of the equation of states of radiation, matter and dark energy for maximally model-independent cosmic history reconstruction.

	\subsection{Prior}
	As prior knowledge, we assume that the signal $s(x)$ exhibits only limited curvature since all known contributions to $s(x)$ like matter, radiation and dark energy contribute linear segments. We therefore punish curvature as measured by $\nabla^2\varphi$ and write our negative log prior as 
\begin{equation}
\mathcal{H}(s-t_\mathrm{bg}) = \frac{1}{2} \int dx\, (s-t_\mathrm{bg})^\dagger \nabla^{\dagger 2} \nabla^2 (s-t_\mathrm{bg}).
\end{equation}
This should control the degree of smoothness of $\varrho(a)=\varrho_\mathrm{crit0}\,e^{s_{x=\ln a}}$ in the following fashion. A one-sigma fluctuation in terms of this prior energy corresponds to a bending of the slope of $s(x)$ by one per e-folding of cosmic expansion. Given that in the standard cosmological model from radiation domination until today the scale factor changed by eight e-foldings (four orders of magnitude) and the combined equation of state of mass in the Universe changed only by four (from radiation to dark energy) we see that that the standard cosmological expansion history is well contained within the one sigma contour of our prior. We can quantify the strength also with respect to data consistency: the variance of $d^2 \ln \varrho(a)/(d\ln a)^2$ is punished with a strength, which requests that for this quantity being on a level of one it is required that the data is under stress by at least one sigma with the reconstructed expansion history.


	
	The signal prior covariance matrix $S$ is then given by 
\begin{equation}
S^{-1} = \nabla^{\dagger 2} \nabla^2,
\label{prior-for-reconstruction}
\end{equation}
which is a diagonal matrix in Fourier space\footnote{The following Fourier convention will be used $$f(k) = \int f(x)\, e^{i kx}\, dx$$. }, 
\begin{equation}
S_{kk'} = 2\pi \delta(k-k') k^{-4}.
\end{equation} 
	
	\subsection{Planck cosmology}
	The $\Lambda$CDM model is based upon a spatially flat and expanding Universe, in which the dynamic of the cosmic expansion is dominated by  the cold dark matter (CDM) and a cosmological constant ($\Lambda$) at late times.
	
	According to the second Friedman equation, the expansion rate $H(a)$ is given by
\begin{equation}
H^2 = \frac{8 \pi G \rho}{3} - \frac{kc^2}{a^2} + \frac{\Lambda c^2}{3},
\end{equation}
where $\rho$ is the energy density, $k$ is the curvature and $G$ is the gravitational constant. 
	
	By defining the density parameters $\Omega_X$ as the ratio between the density of $X$ and the critical density, the second Friedman equation reads 
\begin{equation}
H(a) = H_0 \sqrt{\Omega_\Lambda+\Omega_\mathrm{k} a^{-2}+\Omega_\mathrm{m}a^{-3} + \Omega_\mathrm{rad}a^{-4}}.
\end{equation}
	
	In a flat universe, the curvature density parameter is null, $\Omega_K=0$. Therefore, from the definition of our signal $s(x)=\ln(\rho(x)/\rho_{\mathrm{crit}0})$, the background cosmology is given by
\begin{equation}
t=\ln(\Omega_\Lambda+\Omega_\mathrm{m}a^{-3}+\Omega_\mathrm{rad}a^{-4}) \approx \ln(\Omega_\Lambda+\Omega_m a^{-3}),
\end{equation}
as the $\Omega_\mathrm{rad}\ll 1$ in the late Universe. 
	
	In terms of the coordinate $x=-\ln(a)$ of our signal-space, the former equation is written as
\begin{equation}
t \approx \ln(\Omega_\Lambda+\Omega_m e^{3x}).
\end{equation}
	
	The values we adopt for the density parameters at present time are the ones found by the Planck mission from the cosmic microwave background (CMB) data, $\Omega_m = 0.314$ and $\Omega_\Lambda = 0.686$ \citep{Pla2013}. 
	
	The same values of the $\Omega_m$ and $H_0$ are used for the CDM model. We use this unrealistic CDM scenario to test how much our reconstruction depends on the assumed background cosmology. Since in this case $\Omega<1$, we allow for a curvature term $\Omega_\mathrm{k}=1-\Omega_\mathrm{m}$, which evolve with $a^{-2}$.
	
	\subsection{Agnostic cosmology}
	In order to reconstruct the cosmic expansion history in a way that is agnostic about the assumed constituents of the Universe, like matter, radiation and dark energy, we assume another background cosmology, which we called agnostic cosmology. 
	
	As the $\Lambda$CDM model has contributions to the density evolution of terms proportional to $a^{-4}$ and $a^0$, the agnostic background cosmology is taken as being proportional to $a^{-2}$, which is the geometric mean between these extremes. The exponent $2$ of this background slope is then on log-log scale  
\begin{equation}
\frac{\partial s}{\partial x}=\frac{\partial \ln \rho}{\partial \ln a}= 2.
\end{equation}
Thus, the agnostic background expansion history is given by $t_\mathrm{bg}=2x$.
	
	The prior for the variation around this background expansion is assumed to be the same prior as before, since we do not expect the signal to have any strong curvature. In fact, the agnostic cosmology is the ideal background expansion history for this non-parametric reconstruction,
as log-log density expansion $s(x)$ is not curved and therefore any bending of $s$ and $\varphi$ are exactly equivalent.

	\section{Comparison with other methods}
	In order to emphasize the advantages of our approach, we compare  \texttt{charm} to previous literature. \cite{Ish2011} develop a non-parametric method for the reconstruction of the $H(z)$ based on the principal component analysis (PCA) of the Fisher matrix $F = D^T \Lambda D$. This Fisher matrix corresponds in our notation to the term $R^\dagger N^{-1}R$ in our propagator operator. While \cite{Ish2011} regularize their solution by cutting off higher order PCA components, we can preserve all modes of the expansion, just weight their contribution to the solution according to their bending on log-log scale. Furthermore, our method enforces the positivity of the density $\rho = \rho_\mathrm{crit0}e^s$.
	
	\cite{Arm2006} reconstructs the expansion history with a non-parametric method that also enforces smoothness of the solution. However, 
	in their work the data is smoothed directly to suppress noise (while a background expansion history is temporarily subtracted), while we just encourage smoothness via a prior without modifying the data. In the end, also our approach averages nearby data to suppress noise, but it does this in an adaptive way, having a shorter effective smoothing length in regions of dense data, whereas \cite{Arm2006} employ a constant smoothing kernel in redshift space and have to experiment with the kernel size.
	
	To summarize, our approach requires less tuning of regularization parameters compared to previous approaches, as our prior assumption on the problem solution,  that power law equation of states within a certain range are more natural, already fully specifies the necessary regularization.

	
	\section{Results}
	
	We investigate in this section how far the assumption of a prior cosmology affects the reconstruction. 
	
	First of all, in section 7.1, we show that the iterative Wiener filter is able to successfully reconstruct a perturbation of the $\Lambda$CDM model. For this purpose, a mock data catalog is simulated from a randomly perturbed cosmology. 
	
	Once we have convinced ourselves that our algorithm reconstructs the cosmology from data accurately, we apply it to real data. Now the impact of a prior background cosmology needs to be investigated. For this reason, we start with different expansion histories: the $\Lambda$CDM model, for which we expect that the reconstructed corrections are nearly a constant around zero. Then, we assume the CDM model as our initial guess, for which we expect that the cosmological constant is recovered. Finally we test the agnostic cosmology, in order to see if the iterative Wiener filter is able to reconstruct the standard $\Lambda$CDM cosmology from a non-informative prior.

	\subsection{Mock data}
	Before applying \texttt{charm} to the real data, we validate it with a simulated database. These data were generated from a perturbed Planck cosmology, where the perturbation was a random field.
	
	The aim of generating mock data is to show that our mechanism for the reconstruction of the cosmology is able to find a perturbation of the $\Lambda$CDM model if it is present in the data, even in case the perturbation amplitude is much smaller than the Planck cosmology amplitude. 
	
	The perturbation is introduced as an additive term to the $\Lambda$CDM model, $s = t_{\Lambda\mathrm{CDM}} + \varphi$. It is generated from Gaussian probability density,
\begin{equation}
P(\varphi)\propto \exp\!\left\{\frac{1}{\sigma_A}\int \Big[\frac{1}{2\sigma_c^2} \Big(\frac{\partial^2 \varphi}{\partial x^2}\Big)^2+\frac{1}{2\sigma_\alpha^2} \Big(\frac{\partial \varphi}{\partial x}\Big)^2\Big]  dx \right\},
\end{equation}
that ensures the smoothness of the perturbed field. Here, $\sigma_\alpha$ can be considered as a parameter to control the slope of the perturbation, which we assume to be $\sigma_\alpha = 2$ since this should naturally permit slopes of $\rho \propto a^{-4}$ to $\rho  \propto a^{0}$ bracketing our agnostic background cosmology $\rho \propto a^{-2}$ in case of $\sigma_A=1$. The parameter $\sigma_c$ controls the curvature of the perturbation, which is assume to be 0.5, since we do not expect much curvature, while $\sigma_A$ is a normalization constant to control the amplitude of the perturbation. As we focus on small perturbations, which are the realistic ones, this parameter is fixed to $10^{-1}$. We note that the perturbation is generated from a different probability distribution than the prior used in our inference. This is on purpose to test the robustness of  \texttt{charm}.
	
\begin{figure}[t]
\centering
\includegraphics[width=3.7in]{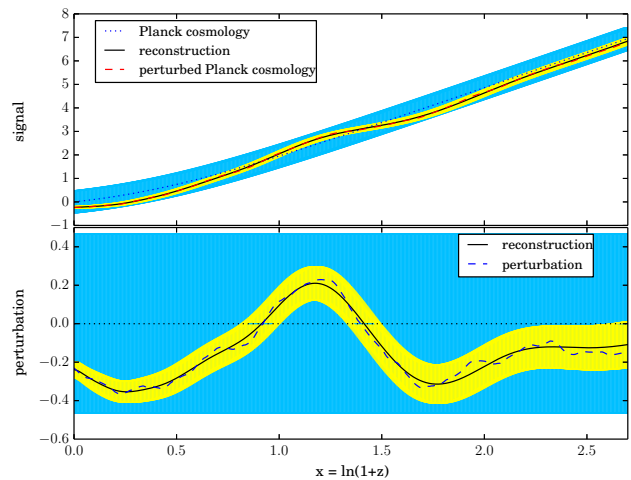}
\caption{Reconstruction using mock data generated with a perturbed Planck cosmology with $\sigma_A=0.1$ (upper panel). The blue and yellow regions correspond to the prior and posterior $1\sigma$ uncertainty limits respectively, which are obtained from the diagonal of the prior and the diagonal of the propagator operator via $\sigma_x = S_{xx}^{1/2}$ and $\sigma_x = D_{xx}^{1/2}$. The bottom panel shows the reconstruction of the perturbation.}
\label{fig:mock}
\end{figure}

		
	Once the perturbed Planck cosmology and mock data are generated, we can apply  \texttt{charm}. The number of simulated SNe Ia is 580 spread in a redshift range of $0.015<z<1.414$, as in the real database. The amplitude of the noise, $\sigma_n$, is set to the amplitude of the perturbation, $\sigma_A$. In the results shown in Figure \ref{fig:mock}, the initial background cosmology is assumed to be the Planck model. It is seen that the perturbation is small, since the perturbated cosmology, used to generate the data, is similar to the Planck cosmology. The prior used for the reconstruction is the one from eq. (\ref{prior-for-reconstruction}) and the result follows the perturbed expansion history. In order to see that the perturbation can be recovered, we plot its reconstruction in the bottom panel of Figure \ref{fig:mock}.
	
	\subsection{Reconstructed cosmology}
	Now, we can apply \texttt{charm} to real data from Union2.1 compilation under the various prior background cosmologies as described before.

\begin{figure}[t]
\centering
\includegraphics[width=3.7in]{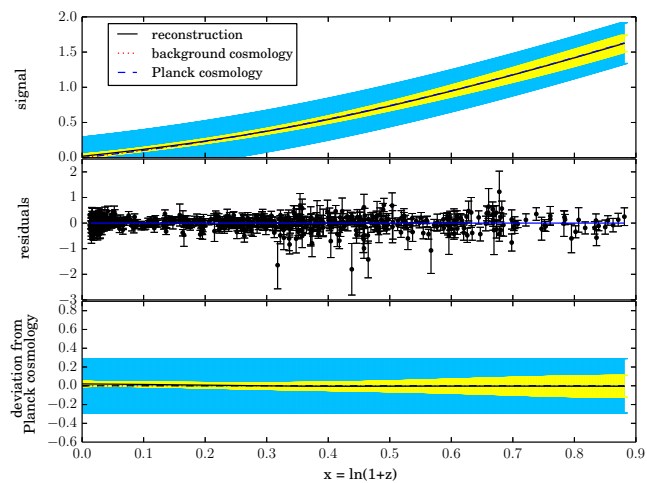}
\caption{Reconstruction assuming Planck cosmology as background cosmology in the upper panel. The blue and yellow regions correspond to the prior and posterior $1-\sigma$ uncertainty limits respectively. The middle panel shows the residuals of the reconstruction and the bottom panel shows the deviation of the reconstruction from Planck cosmology.}
\label{fig:Planck}
\end{figure}
	
	Figure \ref{fig:Planck} displays the results assuming the $\Lambda$CDM model as the cosmology background. The reconstruction is compatible with Planck   $\Lambda$CDM cosmology. The displayed uncertainty limits of the reconstruction are provided by the diagonal of the Hessian, considering all the terms in the inverse propagator $D^{-1}$ plus the term from eq. (\ref{missed-term}) that was was omitted during the numerical minimization of the Hamiltonian for stability reasons. From this, we can conclude that the data agree with the $\Lambda$CDM model, as expected. 
	
	In order to study the agreement of the data with the Planck cosmology, we calculate the signal response, $Rs$, for our reconstruction. The residuals of the real data with respect to this are shown in the bottom panel of Figure \ref{fig:data} and they can be compared to the residuals of the Planck cosmology in the middle panel of Figure \ref{fig:Planck}, calculated by applying the response operator to the Planck cosmology and substracting the real data. We can see that the residuals are almost the same for the Planck cosmology and for our reconstruction.

\begin{figure}[t]
\centering
\includegraphics[width=3.7in]{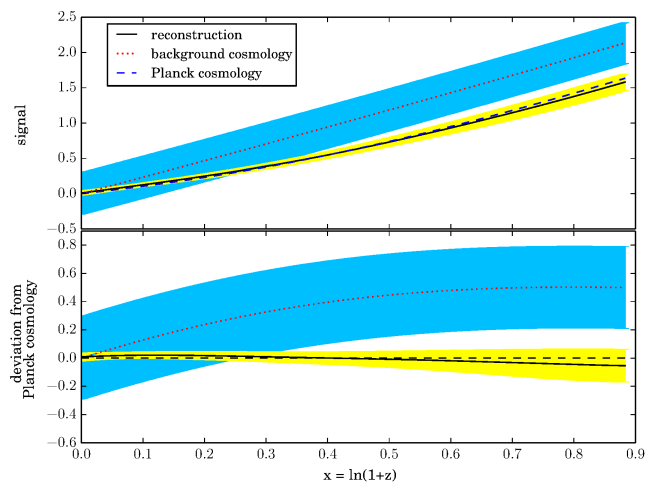}
\caption{Reconstruction assuming CDM model as background cosmology (upper panel) and deviation of the reconstruction from Planck cosmology (bottom panel).}
\label{fig:CDM}
\end{figure}
	
	Figure \ref{fig:CDM} shows the reconstruction for the CDM model as prior background cosmology. 
	Although this model differs significantly from the Planck cosmology, our reconstruction is compatible with the $\Lambda$CDM model. This shows that the data strongly favors a cosmological expansion history dominated by cold dark matter and a cosmological constant. In the following subsection, we recover the density of the cosmological constant $\Omega_\Lambda$ by fitting this reconstruction. The residuals in this case are equivalent to the ones in the middle panel of Figure \ref{fig:Planck}. 
	
\begin{figure}[t]
\centering
\includegraphics[width=3.7in]{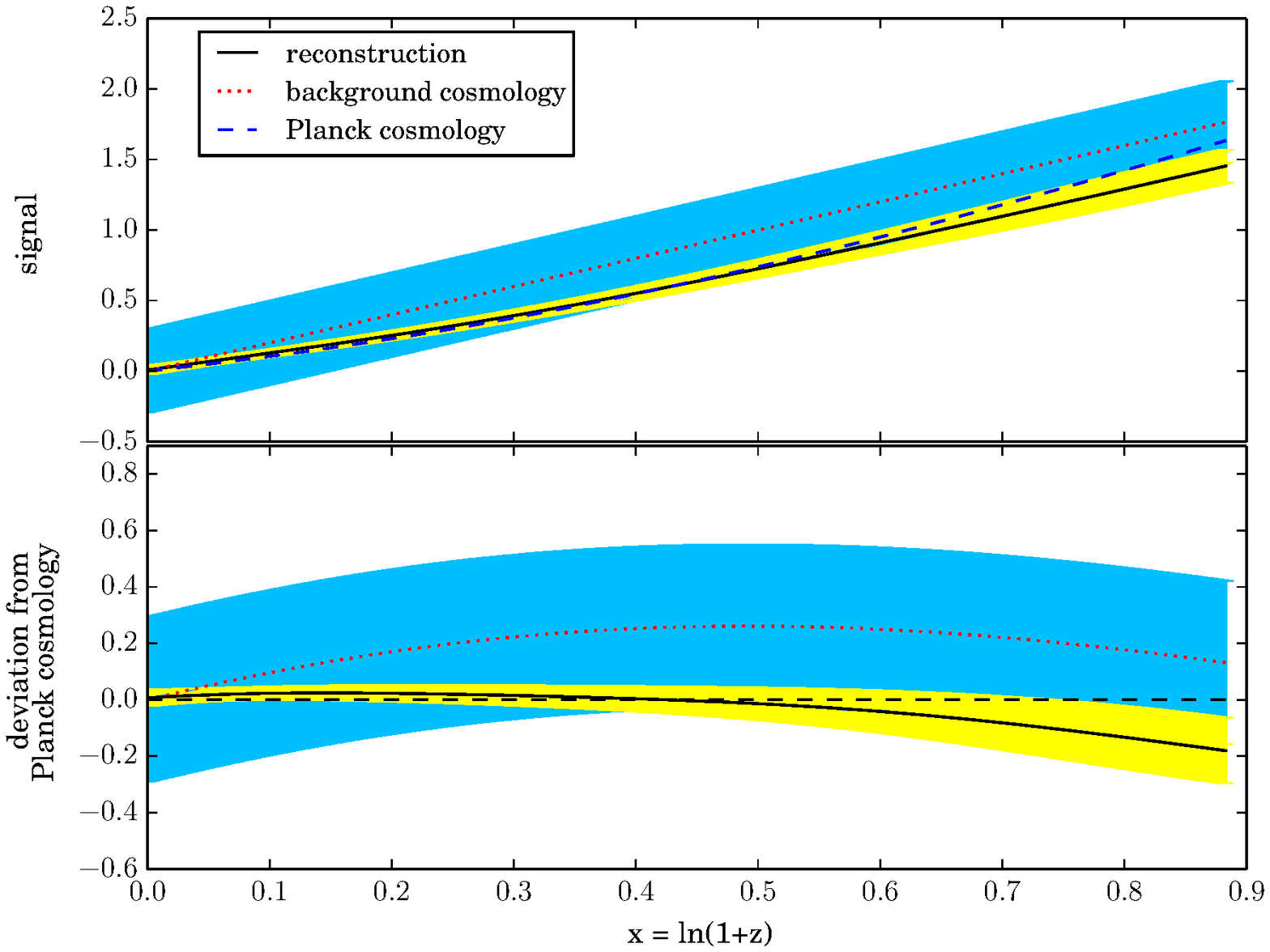}
\caption{Reconstruction assuming agnostic cosmology as background cosmology (upper panel) and deviation of the reconstruction from Planck cosmology (bottom panel).}
\label{fig:agnostic}
\end{figure}
	
	Finally, we adopt the agnostic cosmology as our prior background dcosmology. The result is shown in Figure \ref{fig:agnostic} and demonstrates that the Planck cosmology is recovered even in case  \texttt{charm} is not informed about the specific equations of state of matter, radiation and dark energy. 
	
	From the results in this section, we can conclude that the assumed prior background cosmology does not strongly affect our results. The $\Lambda$CDM is better recovered if it is assumed in the prior, but also a CDM or the agnostic cosmology prior yield a reconstruction close to the Planck $\Lambda$CDM model. Furthermore, we verified by not here displayed experiments that the initial pivot cosmology has no impact on the finally reconstructed cosmology.
	
	\subsection{Fitting $\Omega_\Lambda$ from the reconstruction}
	We have seen in Figure \ref{fig:CDM} that the reconstruction from CDM model recovered the cosmological constant. We can take advantage of this to validate  \texttt{charm} by comparing the density of the cosmological constant that we obtained with the values in the literature. 
	
	In order to obtain the value of $\Omega_\Lambda$ from our reconstruction, we transform our reconstruction in such a way that we can fit a linear regression model. From the definition of the signal as $s = \ln E^2(x)$, this transformation reads as
\begin{eqnarray}
X \equiv e^{3x}, \qquad Y \equiv e^s  \qquad \rightarrow \qquad Y = \Omega_\mathrm{m} X + \Omega_\Lambda
\end{eqnarray}
and thus, the slope of the linear regression corresponds to the mass density and the independent term is identified as the density of dark energy. 
	
	The result of the fitting is $\Omega_\Lambda = 0.670$, which is compatible with the value obtained by Planck mission, $\Omega_\Lambda = 0.686\pm 0.020$ (Planck mission, 2014). This parametric fit to our non-parametric reconstruction should just illustrate the consistency of our reconstruction with other measurements. For a proper uncertainty quantification the parametric model should be fit to the data directly, which was already done by many other works.
	
	\subsection{Constraints and Union3}
	
	\cite{Rub2016} announced recently that a new update of the Union catalog is in preparation, called Union3 Supernova compilation. This new compilation will include high-redshift ($z>1$) supernovae observed with the Hubble Space Telescope. 

\begin{figure}[t]
\centering
\includegraphics[width=3.7in]{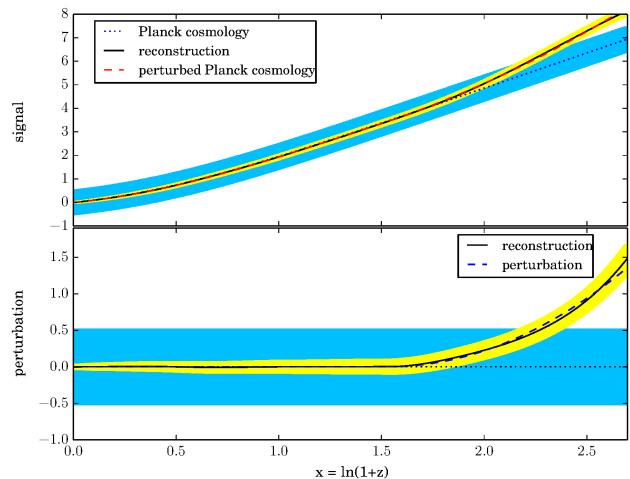}
\caption{Reconstruction of a perturbation of the Planck expansion history at high-redshifts.}
\label{fig:parabola}
\end{figure}

	The study of the BAO provides constraints to the evolution of dark energy density. Recently, BAO measurements have suggested a change in the density parameter of the dark energy at large redshifts $z>1$ \citep{Del2015,  Aub2015}. This early evolution of dark energy could be determined more precisely when data of SNe Ia at high-redshift are available.
	
	In order to prepare our reconstruction algorithm for future work, we check that it is able to reconstruct a perturbation that exists only for large redshifts ($z>1$). We generate a perturbation at $z>2$ by adding a parabola to the Planck cosmology as a test. As in section 7.1, the number of simulated SNe Ia is 580 but now spread in a redshift range of $0.015<z<3.0$, that is an anticipation of the future improved datasets. We can see in Figure \ref{fig:parabola} that this perturbation can be recovered by \texttt{charm}. This reconstruction assumed a Planck model as initial background cosmology and the prior from eq. (\ref{prior-for-reconstruction}). 
	
	
	\section{Conclusions}
	
	In this work we have developed \texttt{charm}, a Bayesian inference method to reconstruct non-parametrically the cosmic expansion history from SNe Ia data and applied it to the Union2.1 dataset. As shown in section 7, the choice of the background prior cosmological expansion history does not significantly affect the final result of the reconstruction. 
	
	We found that the Planck cosmology is reconstructed independently of a prior assumed background cosmological expansion history. We have also recovered the density parameter $\Omega_\Lambda$ in section 7.3 and found a value of $0.670$ which is compatible with the one from \cite{Pla2013}. 
	
	Although no evidence for a deviation from the standard cosmological expansion history has been found, we have tested our method whether it is able to reconstruct a perturbation if it existed. Our method was able to  reconstruct a relative perturbation to $\Lambda$CDM even with an amplitude as low as $10^{-1}$.
	
	Since recent analysis of BAO showed new constraints on the evolution of dark energy density at early epochs \citep{Del2015}, the study of SNe Ia at high-redshift will be crucial in the future for constraining any time variation in dark energy. In order to have \texttt{charm} prepared for the release of the database with supernovae at high-redshift, we have checked that the reconstruction method could find a perturbation that existed only at high-redshifts. Thus,  \texttt{charm} will be a useful tool to study dark energy in a more model-independent way when the new SNe Ia datasets become available.

\begin{acknowledgements}
    Part of this work was supported by Deutscher Akademischer Austauschdienst (DAAD). The computations were done using the NIFTy\footnote{http://ascl.net/1302.013} package for numerical information field theory \citep{Sel2013}
\end{acknowledgements}


\bibliographystyle{aa} 
\bibliography{biblio} 

\begin{appendix}

\section{Systematic errors} \label{appA}
	
	\subsection{Color corrections}
	The empirical color corrections account for dust and intrinsic color-magnitude relation. The validity of the color correction, which is an empirical relation, relies on the assumption that nearby and distant SNe Ia have the same color-magnitude relation. This correction could become a source of systematic error if different corrections were required for different SN populations or if the distance to the SN affected this magnitude-color relation.
	
	The second case seems to be unlikely since the color correction relation at high and low redshift agree. Although this agreement could arise from a different proportion of reddening and intrinsic color at different redshift, it supports the empirical relation for color correction \citep{Kow2008}. 
	
	The presence of two SN populations is supported by two types of SN-progenitors timescales argued by \cite{Man2006}. If this two populations are present, they might evolve in a different way with redshift. If the full sample is divided into equal subsamples by splitting by color, the color correction is significantly different for the two subsamples \citep{Ama2010}. This suggests that the color-magnitude relation could be more complex than a linear relation. 
	
	\subsection{Sample contamination}
	In order to avoid the contamination of the data by non-type Ia SNe, which are not standard candles, an analysis technique was developed by \cite{Kow2008}. This method, which is based on $\chi^2$ minimization, rejects the outliers from the sample. However, the systematic uncertainties are cast into an uncertainty of the absolute magnitude $\Delta M$. In order to consider the sample contamination, an uncertainty $\Delta M= 0.015$ was added to the covariant matrix due to contamination. 
	
	\subsection{Lightcurve model}
	The lightcurve model is a fit with two parameters and it becomes a limit in capturing the diversity of SNe Ia. A problem arises when different techniques are used to observe nearby and distant supernovae, which implies that the parameters are obtained by fitting a different part of the curve in high-redshift SNe. 
	
	A Monte-Carlo simulation was performed by \cite{Kow2008} in order to quantify this systematic error, obtaining that the difference of nearby and distant SNe is $\Delta M =  0.02$ magnitudes. 
	
	\subsection{Photometric peak magnitude}
	The uncertainty of the peak magnitude is due to the color correction. In order to measure the color, the flux is measured in at least two bands. Since the spectra of the SNe at different redshifts are obtained from different bands, their color is determined from different spectral regions. Then, the uncertainty in the different regions of the reference Vega spectrum limits the accuracy of SNe color measurement. 
	
	The Union compilation assumes an uncertainty in the absolute magnitude of $\Delta M = 0.03$ for the photometric peak magnitude. Later, in the Union2 compilation, the numerical effect of each passband on the distance modulus was computed. This was a more efficient way to include this systematic error than including a constant magnitude covariance for all SNe \citep{Ama2010}. 
	
	\subsection{Malmiquist bias}
	Malmiquist bias arises in flux limited surveys. The Union compilation attributes a systematic uncertainty of $\Delta M = 0.02$ in absolute magnitude due to this bias. 
	
	\subsection{Gravitational lensing}
	Gravitational lensing causes dispersion in the Hubble diagram at high redshift \citep{Kow2008}. This effect is treated statistically in the Union compilation. The uncertainty due to gravitational lensing is larger than the intrinsic dispersion only for high-redshift SNe but it causes a bias of magnitudes. This bias is not present if fluxes are used instead of magnitudes. 
	
	\subsection{Galactic extinction}
	The photometry is corrected for galactic extinction using an extinction law that assumes $R_V = 3.1$, together with dust maps \citep{Ama2010}. The galactic extinction is more important in nearby SNe, since the distant ones are measured in redder bands and then, its $R_R$ is approximately the color correction, without an important effect of the galactic extinction. 
	
	\subsection{Host-mass correction coefficient}
	The SNe Ia luminosity is related to the mass of the host galaxy, even after color corrections \citep{Suz2012}. The host galaxies for low-redshift SNe Ia are more massive on average than the host galaxies for high-redshift SNe Ia. This can bias cosmological results and it can be corrected by fitting a step in the mass of the host-galaxy at $m_\mathrm{threshold}=10^{10}M_\odot$. The problem is that for the Union2.1 compilation the individual host galaxy masses are not known. To overcome this problem, a probabilistic method was defined to determine the host mass correction. This procedure could carry systematic errors, that are taken into account in the covariance matrix as $\Delta M = 0.02$.

\section{Simplification of the propagator operator} \label{apb}
	
	
	The propagator operator $D$ measures the convex part of curvature of the information Hamiltonian, since it is the Hamiltonian second derivative except for terms due to higher order non-linearities in the response. This $D$ operator is mostly used to guide our gradient descent via the regularized Newton method. Since the Newton method is not suited for negative curvatures, the term 
\begin{equation}
 \alpha\sum_{ij} \frac{1}{2}\frac{q_{ix}q_{iy}-r_{i}q_{ix}\delta(x-y)}{r_{i}^2} N^{-1}_{ij}\Big(d-R(t)\Big)_j,
\label{missed-term}
\end{equation}
which would be part of the Hessian of the full Hamiltonian is dropped by our response linearisation in order to avoid numerical problems caused by negative Eigenvalues of $D$. This is justified, as $D$ only guides the numerical scheme, while the unmodified $j$ determines where the scheme finally converges to. 
	
	This simplification is allowed because we are iterating the Wiener filter to find the global minimum of the information Hamiltonian, and for this, it is not necessary to perform an optimal step in each iteration just that all steps go in the right direction. Inhibiting negative Eigenvalues of our simplified Hamiltonian ensures that the resulting Newton method always descends toward the global minimum of the Hamiltonian.

\end{appendix}

\end{document}